\documentclass{raa}            

\usepackage{graphicx,times}             
\usepackage{natbib}
\usepackage{amssymb,amsmath}
\bibpunct{}{}{;}{a}{}{,}
\usepackage{hyperref}
\hypersetup{colorlinks = true, linkcolor = blue, anchorcolor = red, citecolor = blue, filecolor = red, pagecolor = red, urlcolor = red}

\usepackage{array}
\usepackage{caption}

\begin{document}

   \title{The Principle of Navigation Constellation Composed of SIGSO Communication Satellites
}

   \volnopage{Vol.0 (200x) No.0, 000--000}      
   \setcounter{page}{1}          

   \author{Hai-Fu Ji
      \inst{1,2}
   \and Li-Hua Ma
      \inst{1}
   \and Guo-Xiang Ai
      \inst{1}
   \and Hu-Li Shi
      \inst{1}
   }

   \institute{National Astronomical Observatories, Chinese Academy of Sciences,
             Beijing 100012, China; {\it mlh@nao.cas.cn}\\
        \and
             University of Chinese Academy of Sciences,
         Beijing 100049, China\\
   }

   \date{Received~~2009 month day; accepted~~2009~~month day}

\abstract{ The Chinese Area Positioning System (CAPS), a navigation system based on GEO
communication satellites, was developed in 2002 by astronomers at Chinese Academy of
Sciences. Extensive positioning experiments of CAPS have been performed since 2005.
On the basis of CAPS, this paper studies the principle of navigation constellation
composed of Slightly Inclined Geostationary Orbit (SIGSO) communication satellites.
SIGSO satellites are derived from end-of-life Geostationary Orbit (GEO) satellites under
inclined orbit operation. Considering the abundant frequency resources of SIGSO satellites,
multi-frequency observations could be conducted to enhance the precision of pseudorange
measurements and ameliorate the positioning performence. The
constellation composed of two GEO satellites and four SIGSO
satellites with
inclination of 5$^\circ$ can provide the most territory of China with 24-hour maximum PDOP less than 42.
With synthetic utilization of the truncated precise (TP) code and physical augmentation factor in
four frequencies, navigation system with this constellation is expected to obtain comparable
positioning performance with that of coarse acquisition code of GPS. When the new approach of
code-carrier phase combinations is adopted, the system has potential to possess commensurate
accuracy of precise code in GPS. Additionally, the copious frequency resources can also be used
to develop new anti-interference techniques and integrate navigation and communication.
\keywords{astrometry and celestial mechanics --- astronomy application --- artificial satellite ---
satellite navigation constellation }
}


   \maketitle

%
%
\section{Introduction}           
\label{sect:intro}

The GPS of America and the GLONASS of Russia could provide Position,
 Velocity and Time (PVT) services for users with global coverage and
all-weather capacity. Chinese Compass has been extending its coverage
 and is scheduled to be a global satellite navigation system around
2020\footnote{http://en.wikipedia.org/wiki/Beidou\_navigation\_system}.
European Galileo is in the initial phase at present and will
launch positioning applications on a global scale as early as
2020\footnote{http://en.wikipedia.org/wiki/Galileo\_\%28satellite\_navigation\%29}.
Several regional navigation systems have also emerged recently, such as the Quasi-Zenith
 Satellite System (QZSS) of Japan, the Indian Regional Navigational
Satellite System (IRNSS). All of the above GPS-like systems require
the launch of specific navigation satellites to constitute constellations.
 In 2002, a group of astronomers at Chinese Academy of Sciences (CAS)
presented a new concept of navigation, to constitute an positioning
 system using Geostationary Orbit (GEO) communication
satellites\footnote{National Astronomical Observatories of China, CAS,
National Time Service Center, CAS, Transponder satellite com-munication
navigation and positioning system, Ai, G. X., Shi, H. L., et al. PRC Patent,
No. 200410046064.1, 2004}, which
is fundamentally different from the GPS-like systems in that navigation
signals are generated on the ground station and then transmitted by
transponders on communication satellites, while receivers decode the
signals and achieve positioning, measurements of velocity and time, and communication.

Under the guidance of this pioneering idea, researchers organized by
CAS successfully developed the Chinese Area Positioning System (CAPS).
Comparing with the GPS-like systems, some advantages are introduced in
CAPS: (1) the cost is much lower and the deployment cycle is notably shorter
 by resorting to available GEO satellites instead of launching specific
navigation satellites; (b) a more accurate and reliable atomic clock can
be employed to provide time reference as navigation signals are generated
 on the ground; (3) inherent communication functions of GEO satellites
 can be embedded in navigation applications to realize the integration
of navigation and communication (\cite{Ai+etal+2008}; \cite{Ai+etal+2009b};
\cite{Shi+etal+2009a}; \cite{Li+Dempster+2010}; \cite{Ma+etal+2012}). In 2005, demonstration
system of CAPS was tested in six cities of China, which successfully passed
 acceptance and adequately displayed the features of CAPS. CAPS was then
included in the guidelines on national medium- and long-term program for
science and technology development (2006-2020) as an important part of the
 Chinese second generation satellite navigation system. Subsequent to the
 successful development of CAPS, the Boeing Company began to make use of
the Iridium communication system to enhance GPS and carry out navigation-related
 applications in 2007\footnote{http://www.coherentnavigation.com/press/2007.04.11\_ee\_times.pdf}.
The Iridium system consists of 66 operational satellites
 positioned in six Sun-synchronous orbital planes. It works in L-band for
 voice data communications and offers much stronger ground receiving signals
 to users than GPS does because of its low-orbit satellites. These features
enable the Iridium system to further GPS.

The constellation with only GEO satellites cannot provide three-dimensional
 (3D) positioning because the satellites are located in coplanar orbit over the equator.
 In CAPS demonstration system, GEO satellites and specific barometric altimeter
were combined to realize altitude-aiding 3D positioning (\cite{Ai+etal+2009a}). In
 this paper, Slightly Inclined Geostationary Orbit (SIGSO) satellites are
recommended to be deployed in navigation system, which could qualify the
system to attain independent 3D positioning. Owing to various perturbations,
orbit elements of GEO satellite vary constantly (\cite{Zhang+1998}; \cite{Ma+etal+2011b}).
 Orbital position of the operational GEO satellite should be maintained at
certain precision. When a GEO satellite approaches the end of its life,
inclined orbit operation is implemented to save propellant fuel and prolong
 its service life, which means the fuel is only consumed for east-west
station-keeping and attitude control. Under inclined orbit operation,
the end-of-life GEO satellite in effect becomes a SIGSO
satellite\footnote{http://www.satsig.net/satellite/inclined-orbit-operation.htm}. SIGSO
satellites have radical improvements on geometric configuration of navigation
 constellation by introducing north-south components and further enable the
navigation system to offer 3D positioning services. Furthermore, by exploiting
plentiful frequency resources of SIGSO satellites, positioning with high
performance could be acquired, as well as the new anti-interference techniques
 and integration of navigation and communication (\cite{Shi+etal+2009};
\cite{Han+etal+2009}; \cite{Cui+etal+2009}; \cite{Ma+etal+2011a}).

The principle of constituting navigation constellation employing SIGSO
 communication satellites is studied in this paper. Perturbations and SIGSO
 communication satellites are introduced in Section 2. We establish
observation equations for the system based on communication satellites
 and perform analyses of Position Dilution of Precision (PDOP) values in
Section 3. Section 4 gives an overview of the navigation performance.
Finally, several issues are further investigated pertaining to the
navigation system based SIGSO satellites.

\section{PERTURBATIONS AND SIGSO COMMUNICATION SATELLITES}
\label{sect:Obs}

\subsection{Perturbations}

The GEO satellite is subject to various perturbing forces, mainly
comprising the Earth nonspherical gravity, lunisolar gravitational
force and solar radiation pressure. Due to the perturbations, motions
of the satellite no longer follow the two-body dynamic equations
 and its inclination, orbital period, eccentricity and Right Ascension
 of the Ascending Node (RAAN) constantly change (\cite{Zhang+1998}; \cite{Ma+etal+2011b}).

Mass distribution of the Earth is uneven and its shape is also irregular.
The polar radius is approximately 23 kilometers shorter than the equatorial
radius. The equatorial plane shows slight oval shape (difference between
semi-major axis and semi-minor axis is about 68 meters). The GEO satellite
consequently gravitates both in tangent and normal directions of the orbit,
and the magnitude depends on the satellite’s position and distance from
the Earth. The longitudes of the minor axis of the equator are at about 75$^\circ$E
and 105$^\circ$W, which have a difference of 90$^\circ$ with that of major axis. The Earth's
 equatorial ellipticity has a long-term accelerating effect on longitude of
the sub-satellite point. As points of minor axis are stable and points of
major axis are instable in geostationary orbit, the equatorial ellipticity
 produces perturbations which cause the satellite to drift back and forth
 in east-west direction around the points of minor axis.

As revolution of the Earth is in ecliptic plane, movements of GEO satellite
 relate not only to gravity acted upon the satellite by the Earth, the Sun
 and the Moon, but also to gravity imposed on the Earth by the Sun and the Moon.
 Lunisolar gravitational force is the most important factor causing the
change of inclination. Because the lunar orbit has an angle between 23.5$^\circ$ $\pm$ 5.1$^\circ$
relative to the equatorial plane, the drift of satellite's inclination varies
 every year and its value depends on the average value of the lunar orbit’s
inclination in a period of 18.6 years. The annual drift rate is from 0.75$^\circ$ to 0.95$^\circ$,
 and its direction alters in different years as the lunar orbit pole changes.
 Since SIGSO satellite has no station-keeping in north-south direction,
the satellite's orbit rotates with a 52-year cycle, centering on a pole
which has an angle of 7.5$^\circ$ with respect to the terrestrial pole. If initial
inclination of a SIGSO satellite is 0$^\circ$, the inclination will reach the
maximum value of about 15$^\circ$ after 26 years and then decrease annually.

The GEO satellite is also exposed to the Sun radiation. One part
of the radiation is absorbed and the other part is reflected. The conversion
of energy involved is known as solar radiation pressure which is the largest
non-gravitational perturbation for the satellite, and hence can have a significant
influence on its orbital dynamics. Solar radiation pressure mainly affects orbital
eccentricity. The movement of eccentricity vector relies on area-mass ratio of the
satellite and the endpoint of the vector forms an ellipse with a period of one year.

\subsection{SIGSO Communication Satellites}

Because of the above perturbations, the GEO communication satellite drifts
ceaselessly. Operational GEO satellite must be kept in a control box in order
to meet the requirements of anti-interference and isolation with adjacent
communication satellites. The GEO satellite therefore always carries propellant
fuel which is consumed to conduct attitude control, and produce east-west
velocity increment against the nonspherical gravity and the solar radiation
pressure, as well as north-south velocity increment against lunisolar perturbation.

Most of the fuel on a GEO communication satellite is expended to maneuver
its position against disturbances in north-south direction. As each satellite
carries a restricted quantity of the fuel, the satellite will come close
to the end of its life after some years in operation. Generally the design
life has certain redundancy, so solar battery and signal transponders in
end-of-life GEO satellite can still serve several years. When attitude control
and east-west maneuvering of an end-of-life GEO satellite are still maintained,
the satellite in fact becomes a SIGSO satellite. Since the
fuel is no longer used for maintaining the orbital position in north-south
direction but only little portion needed for east-west station-keeping and
attitude control, the remaining lifetime of a SIGSO satellite can be extended
largely before being placed into graveyard orbit (\cite{Shi+etal+2009};
\cite{Li+Dempster+2010}; \cite{Ma+2011}). This technique of maneuvering end-of-life GEO
satellites to SIGSO satellites is known as inclined orbit operation.

\begin{figure}[h]
  \centering
   \includegraphics[width=0.8\textwidth]{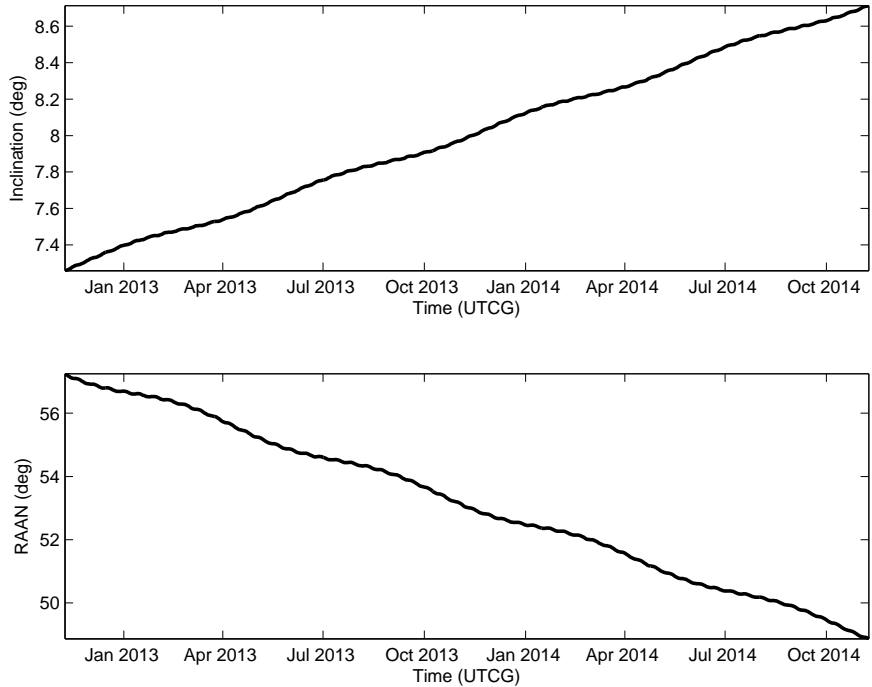}%
   \caption{Orbital evolution of SIGSO satellite under perturbations (\cite{Ma+etal+2011b}).}
   \label{Fig1}
\end{figure}

Evolution of inclination and RAAN of a certain SIGSO satellite is
illustrated in Figure~\ref{Fig1}. Further studies reveal that phase differences
(time differences passing the equatorial plane) of two SIGSO satellites
are determined by their differences in RAAN and differences in longitude
of the sub-satellite point (\cite{Jing+Zhang+2009}). Seeing that SIGSO satellites
follow the same evolution of inclination and RAAN, the phase differences
are mainly determined by the differences in longitude of the sub-satellite point.

\subsection{Advantages and Applications of SIGSO Communication Satellites}

The cost of renting or purchasing end-of-life GEO satellites and then
maneuvering them into SIGSO satellites is obviously lower than that of
developing and launching specific satellites. By deploying SIGSO satellites
in navigation constellation, one can reduce the period and the expense of
constructing navigation system, and make communication satellites play an
important role in navigation and communication applications. As SIGSO
satellites have “figure-8” ground tracks and accordingly can effectively
 improve the geometric configuration by introducing north-south components,
navigation system deploying a certain number of SIGSO satellites can provide
3D positioning services.

Attributable to the plentiful frequency resources of SIGSO satellites,
multi-frequency observations can be performed to achieve centimeter-precision
pseudorange measurement and augment positioning accuracy. Since there are
transponders covering almost whole C band on SIGSO satellites, frequency-switching
techniques can be carried out to reinforce anti-interference performance.
Because each SIGSO communication satellite has frequency band up to 300 MHz or
more, information transmission service based on position, time and status can
be provided aside from positioning service. In addition,
SIGSO satellites can preserve the precious GEO resources (\cite{Shi+etal+2009}).

\section{3. OBSERVATION EQUATIONS AND PDOP ANALYSES FOR THE
    CONSTELLATION BASED ON SIGSO COMMUNICAITON SATELLITES}

\subsection{Observation Equations}

In navigation system based on communication satellites, navigation signals
are generated on the ground station and transmitted via satellite transponders
to users. Using the virtual clock technique, generation time of navigation
signals can be delayed into transmission time at satellite antenna phase
center (\cite{Li+etal+2009a}). To obtain three positions and clock bias, at
least four satellites should be simultaneously observed by users.
Pseudorange observation equation based on code phase from the $i$th satellite
can be modeled as
\begin{equation}\label{equ:eq1}
\rho^i = r^i + I_\rho^i + T_\rho^i + c\cdot\delta t_u + c\cdot\tau_{\mathrm {VCLK}}^i
    + M_\rho^i + \epsilon_\rho^i
\end{equation}
where $i = 1,2,...,M$, and $M$ is the number of observed satellites; $\rho_i$
denotes the pseudorange from antenna phase center on the ground station to
antenna phase center of the receiver; $r^i$ is the geometric distance between the
receiver antenna phase center at signal reception time and the satellite
antenna phase center at signal transmission time; $I_\rho^i$ and $T_\rho^i$ are the ionospheric
and tropospheric propagation delays, respectively; $c$ is the speed of light;
$\delta t_u$ is receiver clock offset; $\tau_{\mathrm {VCLK}}^i$ stands for the
virtual clock parameter, which is broadcasted in the navigation messages;
$M_\rho^i$ and $\epsilon_\rho^i$ account for multipath effects
and measured errors, respectively. We denoteby $\rho_c^i$ the pseudorange obtained
after modifying the virtual clock and propagation delays and compensating
for the multipath effects in the measurements. Equation (\ref{equ:eq1}) is then rewritten as
\begin{equation}\label{equ:eq2}
\rho_c^i = r^i + c\cdot \delta t_u + \epsilon_\rho^i
\end{equation}

To characterize the quality of the positioning estimates, model for the
measurement error in Equation (\ref{equ:eq2}) is simplified as (\cite{Misra+Enge+2006})
\begin{equation}
 E(\epsilon_\rho)=0, Cov(\epsilon_\rho) = \sigma^2 I
\end{equation}
where $E(\cdot)$ represents the mean value, $Cov(\cdot)$ denotes covariance,
$I$ is the identity matrix, and $\sigma$ is the common standard deviation of
 the user range error for each visible satellites. Let $\Delta x$ and $\Delta b$
denote the estimates of positions and clock bias in the local east-north-up (ENU)
coordinate frame respectively, the covariance matrix can been written as
\begin{equation}
Cov \begin{bmatrix}\Delta x \\ \Delta b\end{bmatrix} =
    \sigma^2 \begin{bmatrix}G^T G\end{bmatrix}^{-1} = \sigma^2 H
\end{equation}
where $H=(G^TG)^{-1}$; $G$ is $M\times 4$ matrix with each row composed of three elements of direction
cosine vector represented in ENU coordinate frame, and an entry of 1 in the
last column. The Dilution of Precision (DOP) parameters are defined as (\cite{Misra+Enge+2006})
\begin{subequations}
\begin{align}
\text{Geometric DOP (GDOP)}=\sqrt{H_{11} + H_{22} + H_{33} + H_{44}} \\
\text{Position DOP (PDOP)}=\sqrt{H_{11} + H_{22} + H_{33}}         \\
\text{Horizontal DOP (HDOP)}=\sqrt{H_{11} + H_{22}}                \\
\text{Vertical DOP (VDOP)}=\sqrt{H_{33}}           \\
\text{Time DOP (TDOP)}=\sqrt{H_{44}}
\end{align}
\end{subequations}

The DOP parameters provide simple characterization of the user-satellite geometry.
The positioning error mainly depends upon the measurement geometry and
pseudorange measurement error. The quality of the estimates can be described
as (\cite{Kaplan+Hegarty+2006})
\begin{equation}
\sigma_j = (\mathit{DOP})_j\cdot \sigma
\end{equation}
where $(\mathit{DOP})_j$ designates GDOP, PDOP, HDOP, VDOP or TDOP; $\sigma_j$
is the corresponding estimation errors. It is more common to investigate the PDOP
distributions for the quality of 3D position estimates generally receives extra
attentions from users.

\subsection{PDOP Analyses}

Considering that the navigation coverage of the constellation based on SIGSO
satellites has a north-south symmetry, territories of China are taken for
instance to investigate the constellation configuration. Ten cities in China,
which could be considered as several of representative stations in geographical
distribution, are then selected to analyze PDOP distributions. Daily PDOP
distributions in the ten cities are computed using two GEO satellites situated
in 87.5$^\circ$E and 110.5$^\circ$E and four SIGSO satellites situated in 59$^\circ$E, 71.7$^\circ$E,
125$^\circ$E and 142$^\circ$E respectively. All the six satellites are controlled
by stations within the mainland of China. In the simulation, we presume that
navigation signals can be continuously uplinked to these communication satellites.
Inclination of the four SIGSO satellites is assumed to be 5$^\circ$ and elevation mask
angle is 10$^\circ$. The PDOP values are computed one minute apart over a 24-hour
period and presented in Table~\ref{Tab1} .
\begin{table}
\begin{center}
\caption{Daily Distributions of PDOP for the Ten Cities}
\label{Tab1}
\renewcommand{\arraystretch}{1.5}
\begin{tabular*}{0.75\textwidth}{@{\extracolsep{\fill}} lrrrrr}
\hline

City & Latitude & Longitude & \multicolumn{3}{c}{PDOP}\\\cline{4-6}
& & & max & min & mean \\
\hline
Beijing & 39.90$^\circ$N & 116.47$^\circ$E & 38.3 & 15.3 & 23.1 \\
Shanghai & 31.20$^\circ$N & 121.43$^\circ$E & 37.7 & 15.2 & 22.7 \\
Changchun & 43.89$^\circ$N & 125.31$^\circ$E & 40.9 & 15.5 & 24.7 \\
Taipei & 25.05$^\circ$N & 121.51$^\circ$E & 37.3 & 15.1 & 22.5 \\
Sanya & 18.20$^\circ$N & 109.50$^\circ$E & 36.5 & 14.9 & 22.1 \\
Kunming & 25.05$^\circ$N & 102.70$^\circ$E & 36.8 & 14.9 & 22.3 \\
Lhasa & 29.66$^\circ$N & 91.13$^\circ$E & 37.1 & 15.1 & 22.6 \\
Kashgar & 39.45$^\circ$N & 75.98$^\circ$E & 41.3 & 15.5 & 25.6 \\
Urumchi & 43.77$^\circ$N & 87.60$^\circ$E & 38.4 & 15.5 & 23.3 \\
Xi'an & 34.25$^\circ$N & 108.92$^\circ$E & 37.6 & 15.2 & 22.7 \\
\hline
\end{tabular*}
\end{center}
\end{table}

\begin{figure}[!h]
\centering
\includegraphics[width=0.45\textwidth]{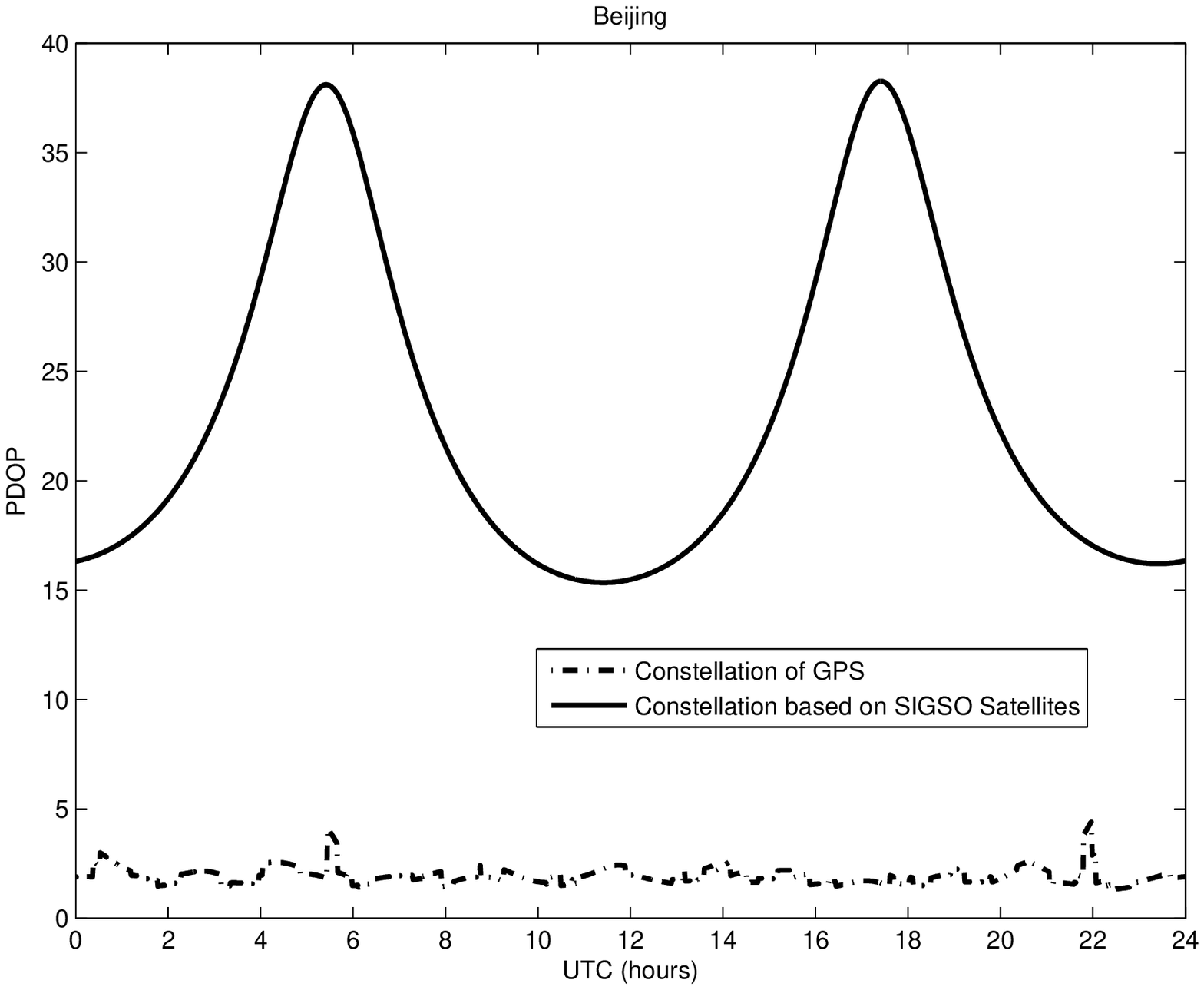}%
\includegraphics[width=0.45\textwidth]{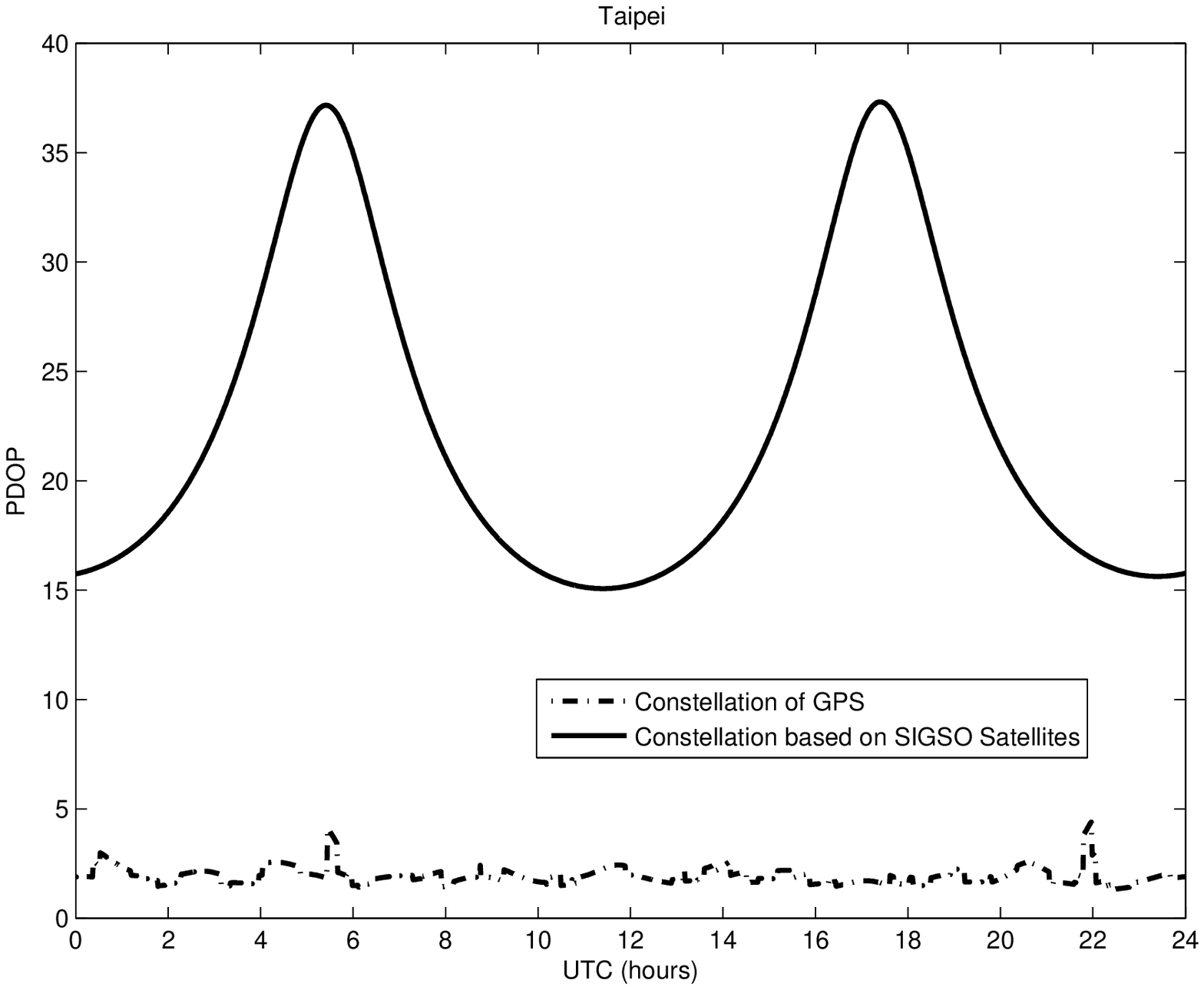}
\includegraphics[width=0.45\textwidth]{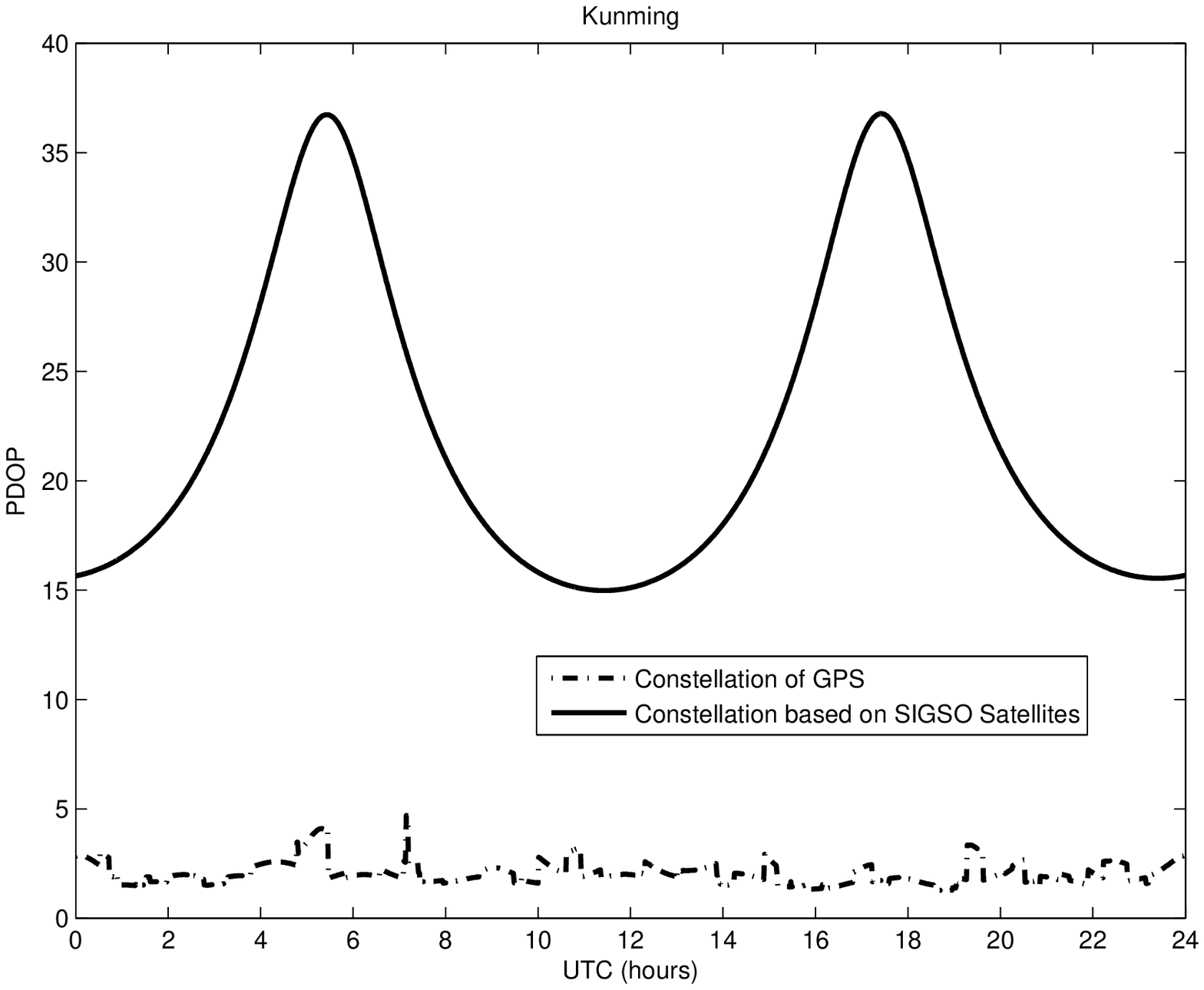}%
\includegraphics[width=0.45\textwidth]{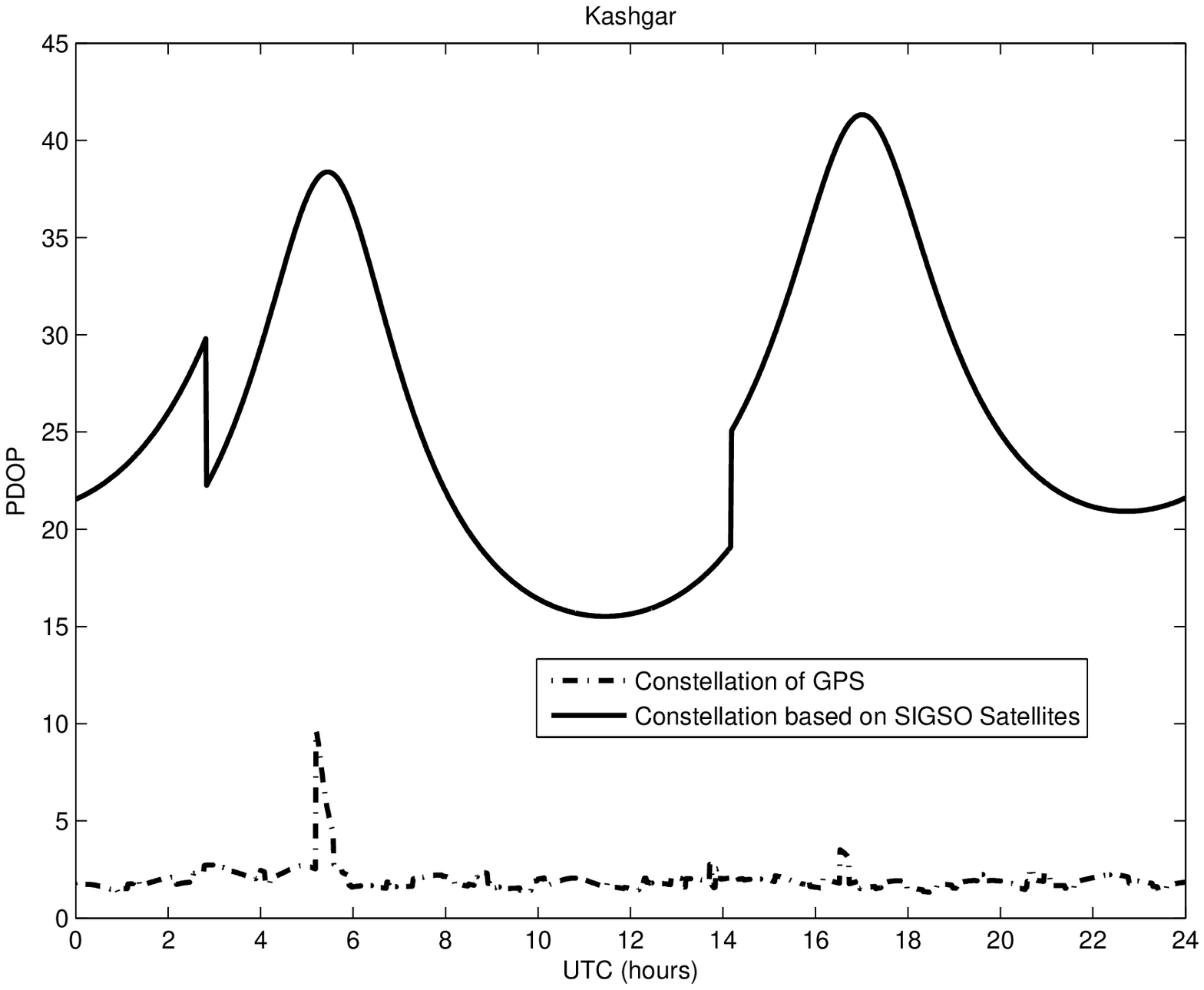}
\caption{Daily distributions of PDOP for the four stations.}
\label{Fig2}
\end{figure}

To give a comparision with GPS, the daily distributions of PDOP for the constellation
of the six satellites and the constellation of GPS in Beijing,
Taipei, Kunming and Kashgar are shown in Figure~\ref{Fig2}.

It can be noted from Table 1 that the maximum PDOP value over 24 hours for
all of the stations is below 42. Moreover, the PDOP values will get
better and better because inclination angles of the SIGSO satellites
increase constantly as time goes on.

\section{AN OVERVIEW OF POSITIONING PERFORMANCE}

\subsection{Multi-frequency observations}
Considering the profuse frequency resources of SIGSO communication satellites,
\cite{Ai+etal+2011} proposed a new approach of code-carrier phase combinations.
In the new approach, three frequencies are selected to structure linear
combinations of code phase and carrier phase. As “Positive” linear combination
and “Negative” linear combination both have ionospheric terms with approximately
same magnitude and opposite sign, integer ambiguity can be fast fixed, and
ionospheric delays can be quickly corrected due to their large differences
between the two selected carrier phase combinations (\cite{Ai+etal+2011}). The new
approach of code-carrier phase combinations possesses the high dynamics of code
phase measurements as well as the high precision of carrier phase measurements.
To further take advantage
of the plentiful frequency resources of SIGSO satellites, navigation signals are
transmitted at multiple frequencies from each satellite and users independently
measure the corresponding multiple pseudoranges, which could enhance the
positioning accuracy significantly. This method is referred to as Physical
Augmentation Factor of Precision (PAFP)\footnote{National Astronomical observatories,
CAS, A method of improving positioning precision in satellite navigation systems,
Ai, G. X., Ma, L. H., Shi, H. L., Ji, H. F., PRC Patent,
No. 201210090864.8, 2012} (\cite{Ai+etal+2012}).

\subsection{Performance evaluation}
The above analyses suggest that positioning accuracy rests with precision
factor (geometric DOP and physical PAFP) and pseudorange measurement accuracy.

\begin{figure}[!h]
\includegraphics[height=0.72\textwidth,angle=270]{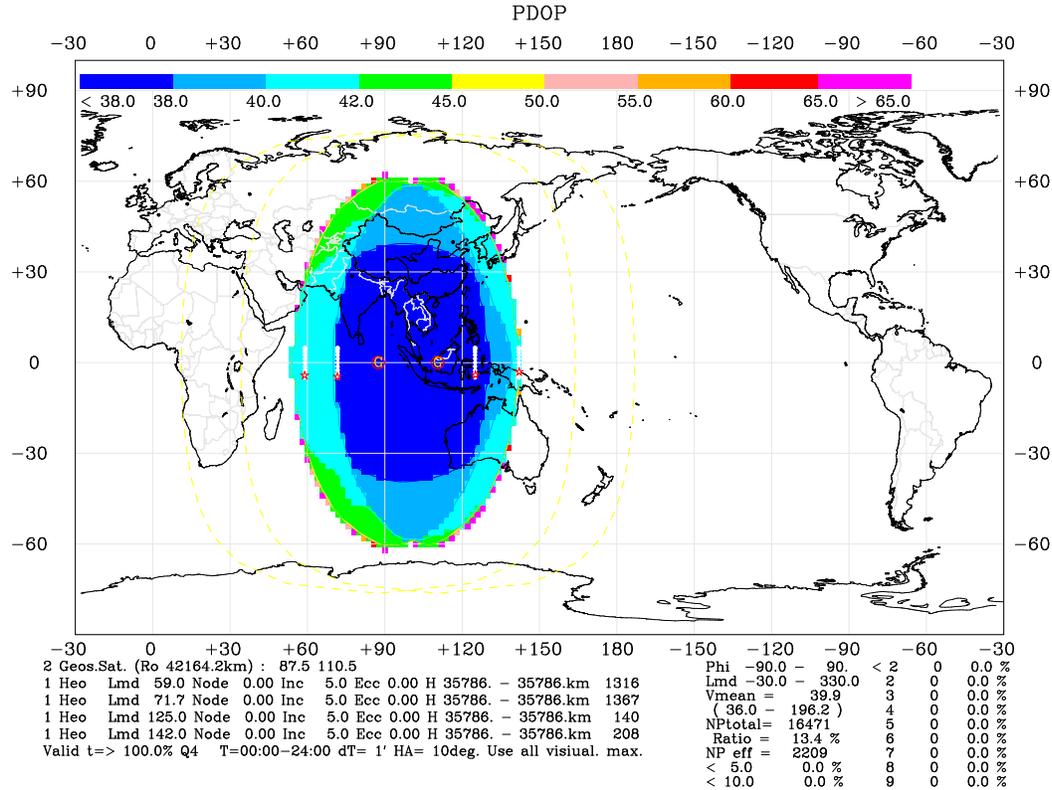}
\setlength{\abovecaptionskip}{-2.8cm} \caption{The global
distribution of maximum PDOP value over 24 hours} \label{Fig3}
\end{figure}

The global distribution of 24-hour maximum PDOP value for the constellation
with the prior six satellites is illustrated in Figure~\ref{Fig3}. As for China, more
than 95 percent of the area can gain a maximum PDOP value smaller than 42
over 24 hours. The constellation provides 92 percent of the area with the
average PDOP value below 25. The minimum PDOP value in 95 percent of the
area is less than 16.

The combined effect of error sources on pseudorange measurements is referred
to as the user equivalent range error (UERE). It is reasonable to model these
errors due to the satellite clock end ephemeris, atmospheric propagation,
multipath, and receiver noise to be uncorrelated, and then the UERE can be
defined as the root-sum-square of these components (\cite{Misra+Enge+2006}). In
transmitted orbit determination method, an accuracy of about 1.0 m of
orbit determination could be obtained (\cite{Li+etal+2009a}; \cite{Li+etal+2009b}).
As the first-order ionospheric effect is inversely proportional to the square
of the carrier frequency, the ionospheric effect in the C-band navigation system
based on SIGSO satellites is much less than in GPS. For tropospheric correction,
it is suggested to establish observation stations on grid points over a large area
to collect atmospheric pressure, temperature and humidity. These actual atmospheric
parameters collected could be used to construct a time-changeable tropospheric
delay model (\cite{Ai+etal+2011}). The rms residual error attributed to atmospheric
propagation models is then assumed to be 0.5 m. The truncated precise (TP)
code with a period of one millisecond and a chip
rate of 10.23 Mcps is adopted in navigation system based on SIGSO communication
satellites. The TP code, having a similar signal structure to GPS precise (P),
is almost ten times more precise than the coarse acquisition (C/A) code in GPS,
so the rms range error due to receiver noise and multipath is assumed to be 0.5 m.
The UERE in the navigation system based on SIGSO satellites is about 1.2 m.

If the TP code is utilized to measure pseudoranges and PAFP in the case of
four frequencies to reduce PDOP twice, the system with the previous constellation
could be expected to have a matching performance with that of GPS C/A code by
synthetic utilization of transmitted orbit determination method, the virtual
clock technique and high-accuracy corrections of tropospheric delays and multipath
effects (\cite{Li+etal+2009a}; \cite{Li+etal+2009b}; \cite{Cheng+etal+2012}). The more the
frequencies used by SIGSO satellites to transmit navigation signals, the greater
the improvement of PDOP values and accordingly the better the positioning
performance. While the new approach of code-carrier phase combinations is
adopted, the system has even capability to provide equivalent positioning
accuracy of P code in GPS.

\section{CONCLUSION}
Due to a variety of perturbations, end-of-life GEO communication satellites
will drift to SIGSO satellites under inclined orbit operation. SIGSO
satellites are favorable to be deployed in navigation constellation to achieve
3D positioning. When SIGSO satellites have relatively large inclination
angles (e.g., 5$^\circ$) and four-frequency PAFP is enforced, the system composed
of SIGSO communication satellites is capable to obtain approximately equal
positioning performance of C/A code in GPS. If the new approach of
code-carrier phase combinations with three frequencies is further employed,
the positioning accuracy will be enhanced remarkably and would be
as good as that of GPS P code.

\section{DISCUSSION}
\subsection{Contributions of IGSO Satellites to Constellation Configuration}
\begin{figure}[!h]
\includegraphics[height=0.72\textwidth,angle=270]{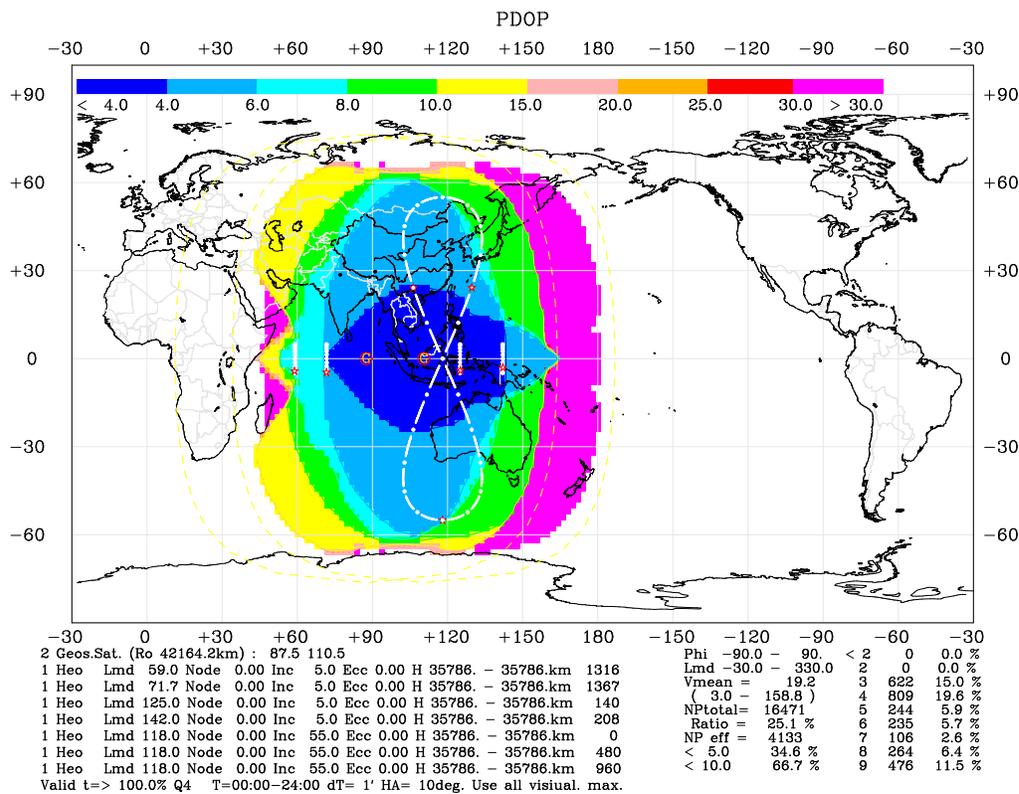}
\setlength{\abovecaptionskip}{-2.8cm} \caption{The global
distribution of maximum PDOP value over 24 hours with three IGSO
satellites.} \label{Fig4}
\end{figure}

In navigation system based on SIGSO communication satellites, if several
IGSO satellites are launched, the configuration will be greatly improved.
One IGSO satellite is not adequate to obtain a satisfying improvement on PDOP
for the reason that users in higher latitudes will fail to receive signals
when the satellite is in the other hemisphere (\cite{Ma+etal+2011a}). We then
adopt three IGSO satellites with inclination of 55$^\circ$ and cross node of
118$^\circ$E for instance, and the previous six satellites to assess the contributions
of IGSO satellites. The amelioration of PDOP values contributed by the three
IGSO satellites is noticeable. The 93 percent of the China area can get a
24-hour maximum PDOP value less than 6.2. The 93 percent of the area obtains
the average PDOP value below 4.8. The minimum PDOP value in the 95 percent
of the area is smaller than 3.5. The global distribution of maximum PDOP
value over 24 hours is shown in Figure~\ref{Fig4}.

However, it should be pointed out that when only four satellites are
observable, the PDOP value is inversely proportional to the volume of a
tetrahedron described by unit vectors from the receiver to each of the
four satellites. If endpoints of the four unit vectors are nearly coplanar,
the PDOP value is infinitely large. For the constellation with three IGSO
satellites and one GEO satellite, there are several time intervals in a
day when the volume of the tetrahedron is close to zero and a 24-hour 3D
positioning is thus not achievable. Further details can be found in
\cite{Ma+etal+2012a}.

\subsection{New Techniques Coping with Interference}
Interference is one of the greatest technical challenges satellite
navigation systems face. The received signal power is lower than natural
noise contained in the bandwidth. If man-made interference is added to
natural noise, then the situation quickly becomes
bleak (\cite{Kaplan+Hegarty+2006}; \cite{Misra+Enge+2006}).

Navigation frequencies are commonly settled when the system has been
designed and in consequence there is no margin for switching frequencies
in applications. As for navigation system based on SIGSO communication
satellites, all C-band frequencies are available for navigation. New
techniques such as frequency-switching, code-switching and satellite-switching
can therefore be employed to enhance anti-interference performance.
Frequency-switching only requires the corresponding change of receiving
frequency when uplink frequency on the ground station alters. In
code-switching technique, the receiver backs up several optional codes
and applies the code adopted in navigation signal to track the signal
and demodulate messages. To utilize satellite-switching technique, the
ground station redirects antennas to other satellites and uplinks
navigation signals, while users receive the transmitted signals to
accomplish positioning.

\subsection{Integration of Navigation and Communication}
Besides PVT services, the transmission of position and time information
is provided in the system based on SIGSO communication satellites.
Satellite link budget, as an elemental optimized design, trades off
various attenuation and gains of transmission links. It involves
system capability and reliability, as well as investment in the
ground station. Navigation transmission link includes both uplink and
downlink, and downlink power is limited. An optimal design is requisite
in order to guarantee successful reception and demodulation of
navigation signals. Communication transmission link deserves more careful
treatments as power is limited in two-way links. In addition,
many system parameters should also conform to the ITU requirements (\cite{Cui+etal+2009}).

\begin{acknowledgements}
The authors are grateful to Profs. Yanben Han and Qiyuan Qiao for
providing DOP analysis software. The work is carried out under the
support of the National Basic Research and Development Program of
China (Grant No. 2007CB815501), the Key Research Program of the
Chinese Academy of Sciences (Grant No. KJCX2-EW-J01) and the
Knowledge Innovation Program of the Chinese Academy of Sciences
(Grant No. KGCX2-EW-407-1).
\end{acknowledgements}

\label{lastpage}

\end{document}